\documentclass[aps,pra,reprint,amsmath,amssymb,showpacs]{revtex4-1}

\usepackage{bm,graphicx,braket}
\newcommand*\ii{\mathrm{i}}

\begin{document}

\title{Relation between $\mathcal{PT}$-symmetry breaking and topologically
  nontrivial phases in the SSH and Kitaev models}
\author{Marcel Klett}
\email{Marcel.Klett@itp1.uni-stuttgart.de}
\author{Holger Cartarius}
\author{Dennis Dast}
\author{J\"org Main}
\author{G\"unter Wunner}

\affiliation{
  Institut f\"ur Theoretische Physik 1,
  Universit\"at Stuttgart,
  70550 Stuttgart, Germany}

\begin{abstract}
  Non-Hermitian systems with $\mathcal{PT}$ symmetry can possess
  purely real eigenvalue spectra. In this work two one-dimensional
  systems with two different topological phases, the topological
  nontrivial Phase (TNP) and the topological trivial phase (TTP)
  combined with $\mathcal{PT}$-symmetric non-Hermitian potentials are
  investigated. The models of choice are the Su-Schrieffer-Heeger
  (SSH) model and the Kitaev chain. The interplay of a spontaneous
  $\mathcal{PT}$-symmetry breaking due to gain and loss with the topological
  phase is different for the two models. The SSH model undergoes a
  $\mathcal{PT}$-symmetry breaking transition in the TNP immediately with the
  presence of a non-vanishing gain and loss strength $\gamma$, whereas the TTP
  exhibits a parameter regime in which a purely real eigenvalue
  spectrum exists. For the Kitaev chain the $\mathcal{PT}$-symmetry
  breaking is independent of the topological phase. We show that the
  topological interesting states -- the edge states -- are the reason
  for the different behaviors of the two models and that the intrinsic
  particle-hole symmetry of the edge states in the Kitaev chain is
  responsible for a conservation of $\mathcal{PT}$ symmetry in the
  TNP.
\end{abstract}

\maketitle

\section{Introduction}
One of the best known relations of topology in solid state systems is
the explanation of the quantized Hall effect, which was first discovered by
von Klitzing et al.\ \cite{Klitzing1980a,Klitzing1986a}, in terms of a
topological invariant \cite{Thouless1982a}. Today topological many-body
systems are a strongly investigated and well understood subject
\cite{Hasan2010a}, and in recent works a topological periodic table has been
proposed~\cite{Altland1997a,Schnyder2008a} to relate topological systems
depending on their symmetries, e.g., electron-particle hole symmetry or
time-reversal symmetry, to different classes. 

Two simple and one-dimensional topological systems are the
Su-Schrieffer-Heeger (SSH) model \cite{Su79a}, initially introduced to
investigate the one-dimensional polyacetylene, and the Kitaev \cite{Kitaev01a}
chain, a model for the description of a one-dimensional spinless
superconductor. They possess an energy spectrum exhibiting a band gap. In
dependence of a certain parameter two different topological phases can arise,
which can be distinguished by energies lying within the band gap. The
corresponding eigenstates of the gap-connecting energies are called edge states.
These edge states show a strong localization at the edge of the system and can
only exist in the topologically nontrivial phase (TNP). Besides the TNP the
two one-dimensional systems feature a topologically trivial phase (TTP), which
is characterized by a fully gapped energy spectrum, in which consequently no
edge states appear.

In reality any topological system will always interact with its nearby
environment, which leads to dissipative effects. A common way to handle
such environment effects in many-body systems is the solution of the dynamics
via Lindblad master equations \cite{Breuer02a}. However, this can become
numerically very expensive, and in many cases an effective description in
terms of the stationary Schr\"odinger equation is sufficient. An often used and
elegant way of describing interactions with an environment on the stationary
level is given by the application of non-Hermitian potentials
\cite{Moiseyev2011a}. Examples range from electromagnetic waves
\cite{Klaiman08a,Wiersig2014a,Bittner2014a,Lawrence2014a,Doppler2016a},
dissipative electric circuits \cite{Stehmann2004a} and optomechanics
\cite{Xu2016a} to quantum mechanics, where it is applied in atomic
\cite{Magunov2001a,Latinne1995,Menke2016a,Cartarius2011b} or molecular
\cite{Lefebvre2009} physics, the scattering of particles
\cite{Hernandez2006,Magunov2003a,Schwarz2015a}, the explanation
of fundamental relations \cite{Heiss1999a,Abt2015a}, and in many-body systems
\cite{Graefe08a,Gao2015a}.

A special class of non-Hermitian operators, viz.\ those possessing a 
parity-time symmetry, has been introduced by Bender and Boettcher in 1998
\cite{Bender98} because these operators feature the interesting property
that they can posses purely real eigenvalues despite their non-Hermiticity.
However, in general the eigenvalues of the non-Hermitian
$\mathcal{PT}$-symmetric operators can be complex. A Hamiltonian is
considered to be $\mathcal{PT}$ symmetric if it commutes with the
combined action of the parity operator $\mathcal{P}$ and the time-reversal
operator $\mathcal{T}$, i.e.\ $[\mathcal{PT},H]=0$. The $\mathcal{PT}$
symmetry of the system can become spontaneously broken, and this symmetry
breaking is related to the realness of the eigenvalues \cite{Bender2007a}. It
can be shown that $\mathcal{PT}$-symmetric eigenstates of a
$\mathcal{PT}$-symmetric Hamiltonian always possess purely real eigenvalues,
while eigenstates that are not $\mathcal{PT}$ symmetric appear in pairs
with complex and complex conjugate eigenvalues. It turned out that
$\mathcal{PT}$ symmetry is a powerful concept to effectively describe systems
interacting with an environment in such a way that they experience balanced gain
and loss. In particular, it was shown in optical experiments that $\mathcal{PT}$
symmetry and $\mathcal{PT}$-symmetry breaking can be realized in the laboratory
\cite{Rueter10a,Peng2014a,Guo09a}. Proposals for the realization in
quantum mechanics exist for Bose-Einstein condensates
\cite{Greafe2012a,Kreibich2016a}.

Recently some models of topological insulators have been investigated under
gain and loss effects in terms of non-Hermitian operators. This leads to
new questions. In particular, it has to be understood whether topologically
protected states can be found in presence of the gain and loss
\cite{Hu2011a,Esaki2011a,Schomerus2013a,Zeuner2015a,Yuce2015a,Yuce2015b,%
Yuce2016a,SanJose2016a}. In an optical experiment of a modified SSH model
topological interface states were observed \cite{Weimann2016a}. Even though
the SSH and Kitaev models are equivalent in some special cases
\cite{Cobanera2015a} they behave completely differently when complex
on-site potentials are applied. Zhu et al.~\cite{Zhu14a} and Wang et
al.~\cite{Wang15} have studied the connection between the TNP and spontaneous
$\mathcal{PT}$-symmetry breaking due to external gain and loss in the SSH and
Kitaev models, respectively. Comparing the results of the two investigations
leads to a discrepancy in the interplay between topological phases and
spontaneous $\mathcal{PT}$-symmetry breaking. In the Kitaev chain the
$\mathcal{PT}$ symmetry is protected within the TNP when a non-Hermitian
potential is applied. On the other hand the SSH model shows an instantaneous
$\mathcal{PT}$-symmetry breaking within the TNP for every arbitrarily small
gain and loss effect. Also in further models it was sometimes found that
completely real eigenspectra do not appear in the TNP, whereas this was
possible in other models.

It is the purpose of this paper to give an unambiguous answer to the question
of how the relation between topologically nontrivial edge states and the effects
of $\mathcal{PT}$-symmetry breaking can be established. To do so, we
investigate the SSH and Kitaev models in the presence of two different
non-Hermitian potentials generating $\mathcal{PT}$-symmetric gain-loss
effects. In particular we study the eigenstates of the system and the
symmetries of the edge states. It will turn out that there is no general
relation between the $\mathcal{PT}$ symmetry of the system and the topological
phase as assumed previously \cite{Wang15}. The symmetry of the specific edge
states in the systems decides whether or not these states spontaneously break
the $\mathcal{PT}$ symmetry. The symmetry the states exhibit in the Hermitian
case survives in the presence of the gain-loss effect. However, in dependence
of the imaginary potential applied to the system also the bulk states may lead
to a spontaneous $\mathcal{PT}$-symmetry breaking in both the TNP and the TTP.

The paper is organized as follows. In Sec.\ \ref{sec:models} the two
different Hamiltonians are introduced. In Sec.\
\ref{sec:results_and_discussions} energy spectra of the two models are shown
without and with the application of external gain and loss potentials. This
is used to analyze the cause of $\mathcal{PT}$-symmetry breaking in the
TTP and the TNP. In particular, the different symmetry behaviors of the
topologically interesting edge states are presented. For the Kitaev chain the
number of edge states is counted for certain parameter values to investigate
their dependence on the imaginary potentials. The last Sec.\ \ref{sec:summary}
provides conclusions.

\section{The Models}
\label{sec:models}

In this paper we consider two different one-dimensional models with a
lattice distance $a=1$ and $N$ lattice sites. The Su-Schrieffer-Heeger
model \cite{Su79a} is given by
\begin{equation}
  \label{eq:ssh-hamilton}
  H_{\textrm{SSH}} = \sum_{n} \left ( t_- c^\dagger_{2n-1}c_{2n} 
    + t_+ c^\dagger_{2n} c_{2n+1} + \textrm{h.c.} \right )  \; ,
\end{equation}
where the alternating hopping strengths $t_{\pm} = t(1\pm \Delta \cos \Theta)$
contain the hopping amplitude $t$ and the dimerization strength $\Delta \cos
\Theta$, which can vary from $-\Delta$ to $\Delta$. The second system is the
one-dimensional Kitaev chain~\cite{Kitaev01a}, which is a toy model for a
topological $p$-wave superconductor,
\begin{equation}
  \label{eq:kitaev-hamiltonian}
  H_{\textrm{Ki}} = \mu \sum_{n} c^\dagger_n c_n + \sum_{n} (t c^\dagger_n c_{n+1} -\ii \delta c_n c_{n+1} + \textrm{h.c.}) \; ,
\end{equation}
where the chemical potential is given by $\mu$, $t$ is again the
nearest neighbor hopping and $\delta$ is the $p$-wave pairing
amplitude. In both models the operator $c_n^\dagger$ ($c_n$) is the
creation (annihilation) operator for electrons at lattice site $n$.
In the following all energies are measured in units of $t$, i.e.,
$t = 1$ is always set, which defines the dimensionless units used in this
work.

In our study the two systems are described by the total Hamiltonians
\begin{equation}
  \label{eq:total-hamiltonians}
  H = H_0 + U \; ,
\end{equation}
where $H_0$ is either the Hamiltonian of the Kitaev model
$H_{\textrm{Ki}}$ or the SSH model $H_{\textrm{SSH}}$. The term $U$
represents the gain and loss effects via an additional
$\mathcal{PT}$-symmetric part. In this work we distinguish between two
potentials,
\begin{equation}
  U_1 = \ii \gamma c^\dagger_1 c_1 - \ii \gamma c^\dagger_N c_N \; ,
  \label{eq:u1}
\end{equation}
in which electrons gain in probability amplitude at the first site and
lose at the last site. The second $\mathcal{PT}$-symmetric potential,
\begin{equation}
    U_2 = \ii \gamma \sum_{n} (-1)^n c^\dagger_n c_n \; ,
    \label{eq:u2}
\end{equation}
corresponds to an alternating gain and loss effect of the whole chain.

Due to the superconducting term in the Kitaev
Hamiltonian~\eqref{eq:kitaev-hamiltonian} a coupling between electrons
and holes arises. The basis of the Kitaev chain has to be expanded to
respect the particle-hole coupling. The particle number operator of an
electron at site $i$ is given by the relation $n_{\mathrm{e},i} = c_i^\dagger
c_i$, whereas the number operator for holes reads $n_{\mathrm{h},i} = c_i
c_i^\dagger$. A matrix representation of the
Hamiltonian~\eqref{eq:kitaev-hamiltonian} can be achieved by choosing vectors in
the form of 
\begin{equation}
  \label{eq:basis-choice-kitaev}
  \ket{\psi} = (\bm{c}, \bm{c}^\dagger )^T
\end{equation}
with $\bm{c} = (c_1 ,c_2, \ldots c_n)$ and $\bm{c}^\dagger =
(c^\dagger_1 ,c^\dagger_2, \ldots c^\dagger_n)$. The projection
$\braket{\psi|\psi}$ corresponds to all number operators of electrons
and holes.

\section{Energy spectra and  phase diagrams}
\label{sec:results_and_discussions}

In this section the numerical solutions of the single-particle
eigenvalue equation
\begin{equation*}
  H \ket{\psi} = E \ket{\psi}
\end{equation*}
are calculated under open boundary conditions (OBC). The Hamiltonian
is given by $H= H_0 + U$, where $H_0$ is the Hamiltonian of the
considered model and $U$ is one of the two $\mathcal{PT}$-symmetric
potentials $U_1$ or $U_2$. Due to the $\mathcal{PT}$ symmetry of the
Hamiltonian solving the eigenvalue equation for an applied gain and
loss effect can lead to purely real eigenvalues, however, in general the
eigenvalues are complex numbers $E = \mathcal{E} + \ii \Gamma$ with the
real energy part $\mathcal{E}$ and the decay or growth rate $\Gamma$.

\subsection{Hermitian system}
\begin{figure}[tbp]
  \centering
  \includegraphics[width=0.5\textwidth]{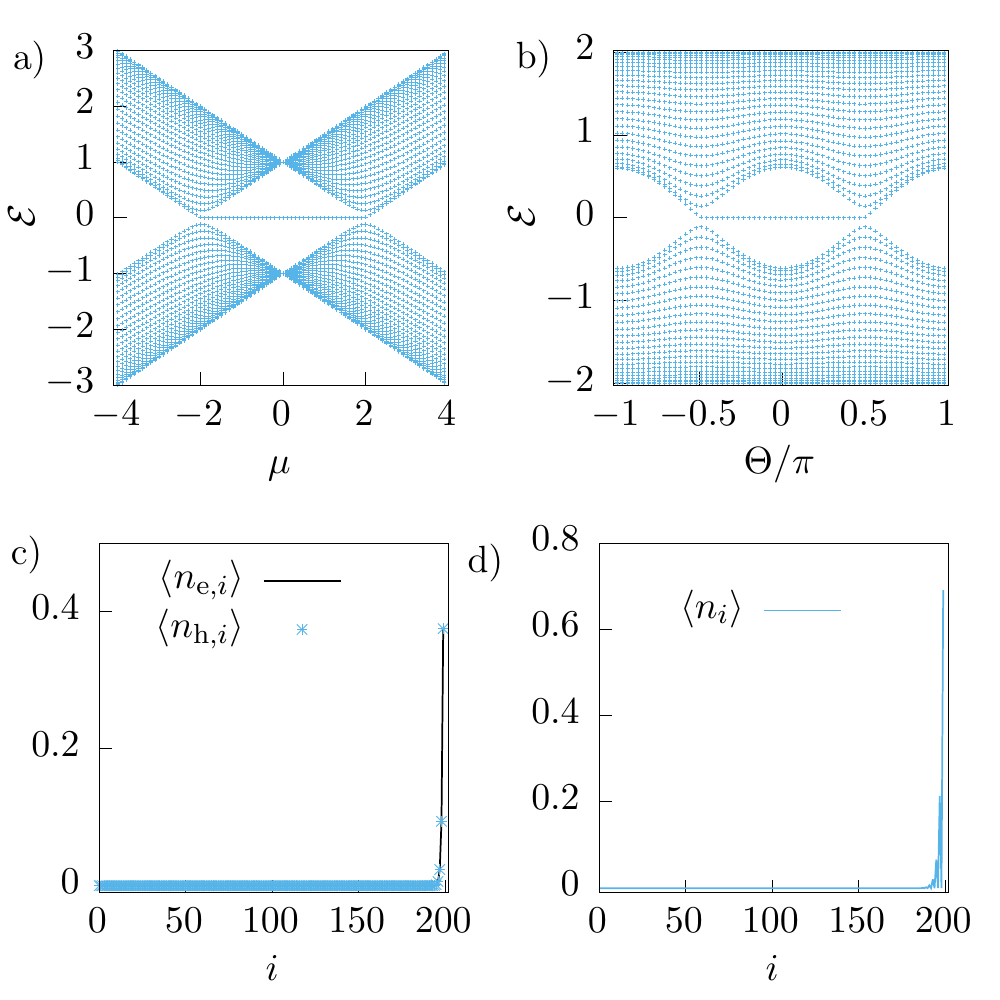}
  \caption{First row: isolated ($\gamma=0$) energy
    spectrum for a) the Kitaev chain with the parameters $t=1.0$,
    $\delta = 1.0$ and b) the SSH chain with $t=1.0$, $\Delta=0.3$. In
    both cases the spectrum is computed for a chain with $N=200$
    sites. Second row: One of the two edge states for c) the Kitaev
    model with the parameters $\mu=1.0$, $t=1.0$ and $\delta = 1.0$
    and d) for the SSH model with $\Theta = 0.1\pi$, $t=1.0$ and
    $\Delta=0.3$. The edge states shown feature a vanishing energy
    eigenvalue.}
  \label{fig:energy-isolated}
\end{figure}

For the reader's convenience we briefly recapitulate the essential properties
of both models. In the case of the isolated models, i.e., $H=H_0$, both energy
spectra show domains, in which a vanishing energy emerges. The presence of a
zero-energy is connected to edge states. The parameter regime hosting
edge states belongs to the topological nontrivial phase (TNP). This phase is
called topologically nontrivial since a topological invariant can be found.
The edge states differ in this invariant from the bulk states
 \cite{Takayama80a, Su80a,Heeger88a,Kitaev01a,Alicea12a}.
In the Kitaev chain the topologically nontrivial phase ranges from $\mu = -2
\ldots 2$, whereas the TNP reaches from $\Theta = -\pi/2 \ldots \pi/2$ in
the SSH model, see figure~\ref{fig:energy-isolated} for $N\to \infty$ lattice
sites.
The calculation of the expectation value of each particle number
operator at every lattice site can be used to illustrate the
localization of the electrons in a certain eigenstate along the
chain. The states of interest are the edge states. For the SSH model the
required expectation value of the occupation of lattice site $i$ is
calculated via
\begin{equation}
  \label{eq:ssh-expectation}
  \braket{n_i} = \braket{\psi_{\textrm{ed}}| c^\dagger_i c_i | \psi_{\textrm{ed}}} 
  \; ,
\end{equation}
where $\ket{\psi_\textrm{ed}}$ is one of the two edge states with zero
energy. For the Kitaev chain one has to distinguish between the
expectation values of particles and holes,
\begin{equation}
  \braket{n_{\mathrm{e},i}} = \braket{\psi_{\textrm{ed}}| c^\dagger_i c_i | 
    \psi_{\textrm{ed}}} \; ,  \quad
  \braket{n_{\mathrm{h},i}} = \braket{\psi_{\textrm{ed}}| c_i c^\dagger_i | 
    \psi_{\textrm{ed}}} \; .
\end{equation}
In figure~\ref{fig:energy-isolated} one edge state for each model is
shown. As one can see the edge state of the Kitaev model fulfills
particle-hole symmetry, i.e., $\braket{n_{\mathrm{e},i}} = \braket{n_{\mathrm{h},i}}$.
Even though no potential is applied the $\mathcal{PT}$ symmetry of the
edge states shown can be explored. In the SSH model the $\mathcal{PT}$
symmetry is broken by the edge states, whereas the edge states of the
Kitaev model conserve the $\mathcal{PT}$ symmetry. Though every single
expectation value $\braket{n_\mathrm{e}}$, respectively $\braket{n_\mathrm{h}}$,
of the Kitaev edge state is not $\mathcal{PT}$ symmetric, it is the
particle-hole symmetry, which ensures the $\mathcal{PT}$ symmetry of
the edge state. Due to the anticommutation relation of Fermions
$\{c_i, c_j^{\dagger} \} = \delta_{i,j}$ a gain $\gamma$ of an electron
at site $i$ corresponds to the equal loss of a hole at the same
lattice site. Applying a $\mathcal{PT}$-symmetric potential generates
gain and loss effects to specific lattice sites in the Kitaev chain
depending on the potential $U_1$, respectively $U_2$. For one lattice
site the net gain-loss effect is zero if a particle-hole symmetry in
the occupation probabilities is present. If an eigenstate preserves
particle-hole symmetry throughout the whole system, the net gain is
zero and therefore the corresponding eigenstate accomplishes
$\mathcal{PT}$ symmetry. The crucial question now is whether this symmetry
survives in the case that imaginary potentials are indeed applied.

\subsection{Small gain and loss effects}

For small gain and loss strengths $\gamma$ the spectra of the energy
real parts do not change much under variation of the imaginary potential
as compared to the isolated cases. Moreover the parameter regimes of
the TNP stay the same in the Kitaev model as well as in the SSH model,
and therefore also edge states are available in the case of small
imaginary potential strengths. The calculation of the edge state
expectation values generate the same localization as in the isolated
case. Thus the total situation does not change and the edge states
remain $\mathcal{PT}$ symmetric in the Kitaev chain, whereas both edge
states of the SSH model are $\mathcal{PT}$-broken eigenstates. 

\begin{figure}[tbp]
  \centering
  \includegraphics[width=0.5\textwidth]{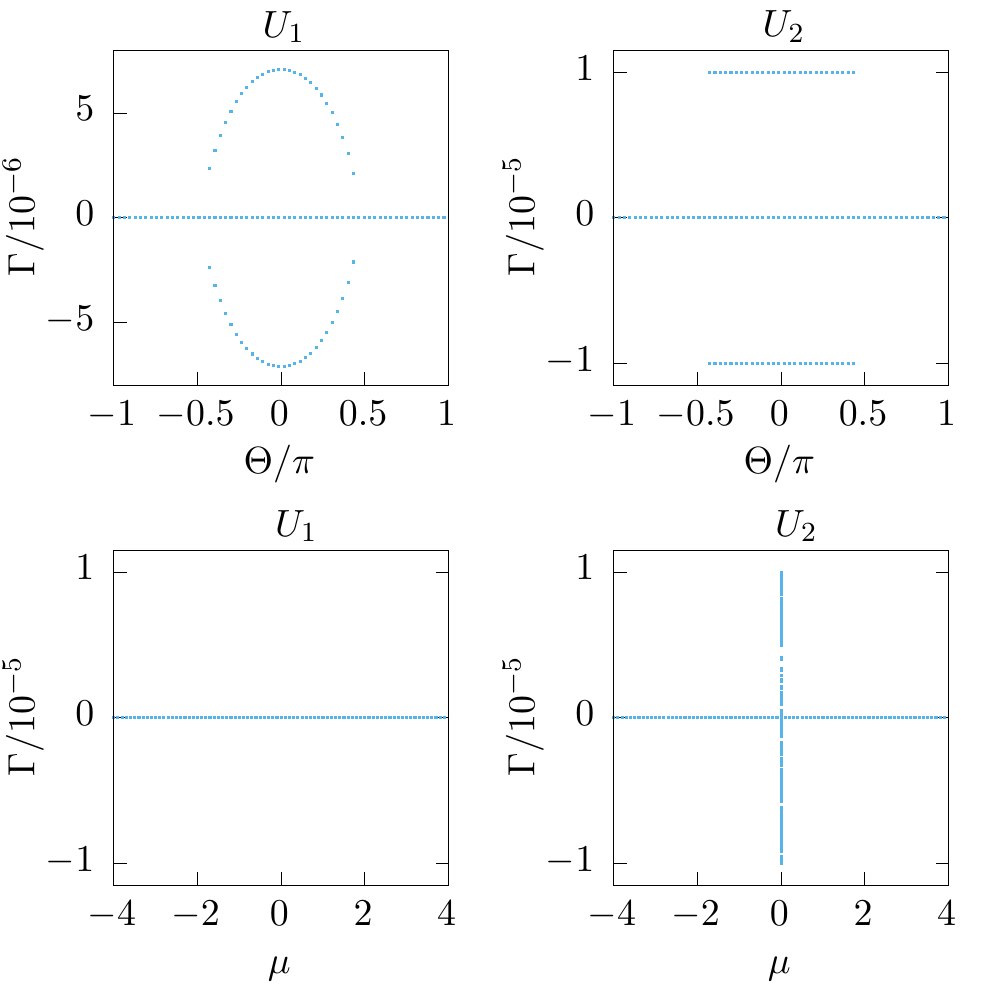}
  \caption{First row: Imaginary parts of all energies of
    the SSH model with potentials $U_1$ (left) and $U_2$ (right). In
    both calculations the parameters $t=1.0$, $\Delta=0.3$, $N=200$, and
    $\gamma = 10^{-5}$ were used. Second row: Imaginary parts of
    all energies of the Kitaev chain with the parameters $t=1.0$,
    $\delta=1.0$, $N=200$, and $\gamma = 10^{-5}$. }
  \label{fig:small-strength}
\end{figure}

A purely real energy spectrum can only occur if every eigenstate of
the system obeys $\mathcal{PT}$ symmetry. Due to the fact that
even for any arbitrarily small gain and loss effect at least the
two edge states of the SSH model break $\mathcal{PT}$ symmetry
the energy spectrum has to show complex energies in the TNP. This is indeed
the case which is illustrated in the first row of
figure~\ref{fig:small-strength}.
For both potentials two complex energies emerge in the case of the SSH
model. The spectrum performs a $\mathcal{PT}$ phase transition at the
same parameter, at which a topological phase transition occurs in the
isolated case, i.e., at $\Theta = \pm \pi/2$.

In the second row of figure~\ref{fig:small-strength} the imaginary parts
of the energy spectrum for the Kitaev chain with applied potentials
$U_1$ and $U_2$ are shown. In contrast to the SSH model the potential
$U_1$ preserves a purely real spectrum. Both topological phases show
the same behavior related to the $\mathcal{PT}$ symmetry, and
therefore the imaginary parts of the eigenvalues cannot be used to
distinguish between the TNP and TTP. For the Kitaev model with applied
potential $U_2$ the imaginary part of the energy shows a non-vanishing
value for disappearing chemical potential $\mu =0$. Still the energy
spectrum of the imaginary part cannot be used to provide any
information about the topological phases. Taking a closer look at the
eigenstates corresponding to the violation of the $\mathcal{PT}$
symmetry for $\mu=0$ reveals the fact that all states with complex
energies are bulk states.

The important finding is that the properties of the edge states in both models
are not altered immediately by the presence of the gain-loss effect. In
particular, the symmetries of the edge states survive, which leads to an
immediate $\mathcal{PT}$-symmetry breaking in the SSH model by the edge states
and a preserved $\mathcal{PT}$ symmetry in the Kitaev chain. This explains
that it is the symmetry of the actual edge states that is connected to the
$\mathcal{PT}$ symmetry of the whole system not the existence of a
topologically nontrivial phase alone.

\subsection{$\mathcal{PT}$-symmetry breaking in dependence of
  $\gamma$}

\begin{figure*}[tb]
  \centering
  \includegraphics[width=0.95\textwidth]{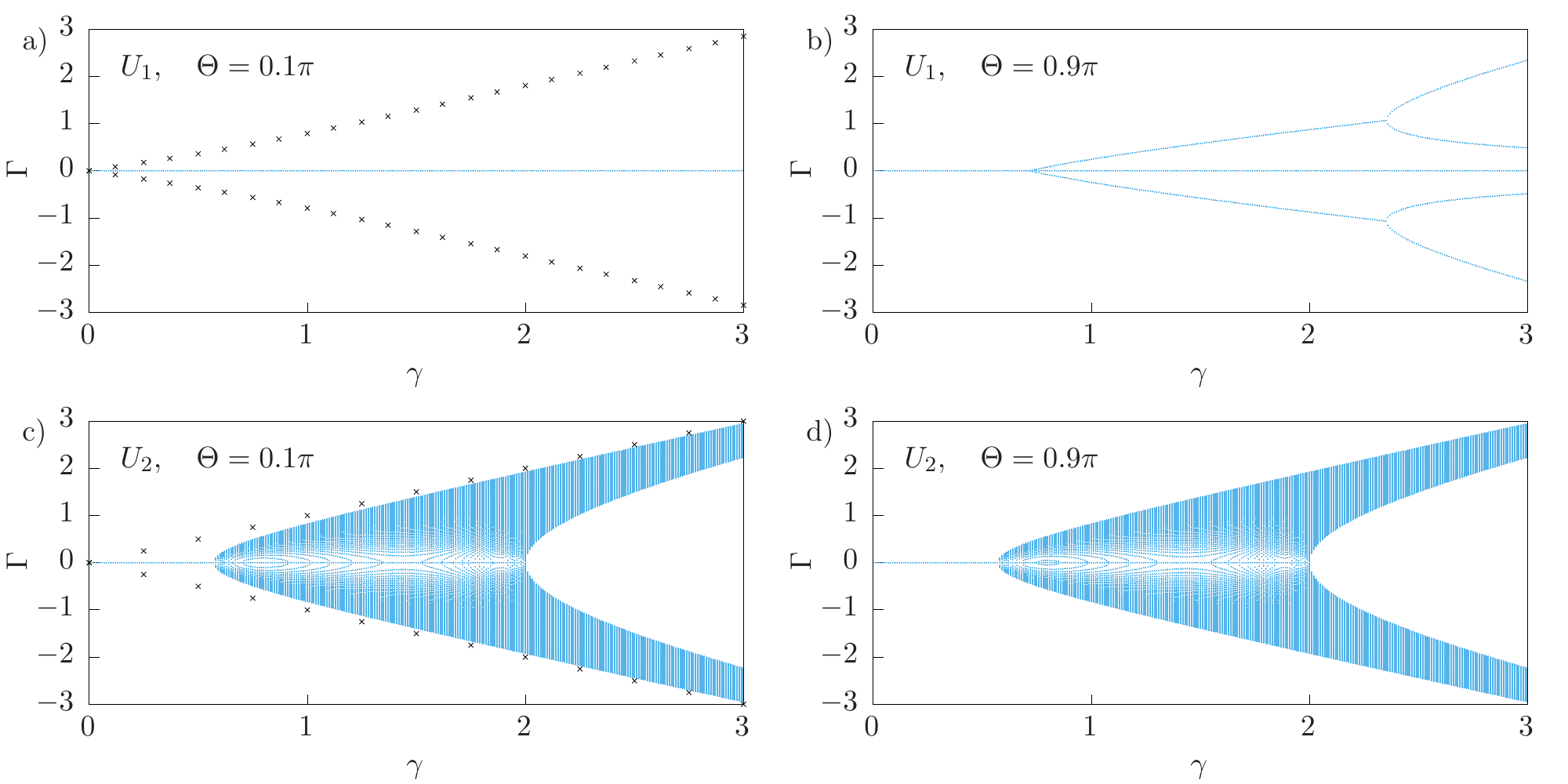}
  \caption{Imaginary parts of all energies in dependence
    on the gain and loss strength of all eigenstates of the SSH
    model. For each plot the parameters $t=1.0$, $\Delta = 0.3$ and
    $N=200$ are used. For a) and b) the potential $U_1$ is used,
    whereas for c) and d) the potential $U_2$ is applied. The black
    crosses in a) and c) represent the purely imaginary energies of the two
    edge states appearing in the TNP.}
  \label{fig:ssh-pt-phasediagram}
\end{figure*}

Leaving the field of a low gain and loss effect a $\mathcal{PT}$
phase diagram can be realized by plotting the imaginary parts of the
energies over the potential strength $\gamma$. In the case of the SSH
model the phase diagrams for both the potentials $U_1$ and $U_2$ are
shown in figure~\ref{fig:ssh-pt-phasediagram}.
For the dimerization strength $\Theta = 0.1\pi$, i.e., in the TNP, the
potential $U_1$ only shows one pair of complex conjugate eigenvalues, which
vanish for $\Theta = 0.9\pi$ in the TTP. The same pair also appears if the
potential $U_2$ is applied. The corresponding eigenstates for the two complex
eigenvalues appearing in both potentials are the two existing edge states. In
contrast to the potential $U_1$ both of the dimerization strengths shown for
$U_2$ possess a critical value, at which the system gets completely
$\mathcal{PT}$ broken and therefore not a single energy eigenvalue remains
purely real.

\begin{figure*}[tb]
  \centering
  \includegraphics[width=0.95\textwidth]{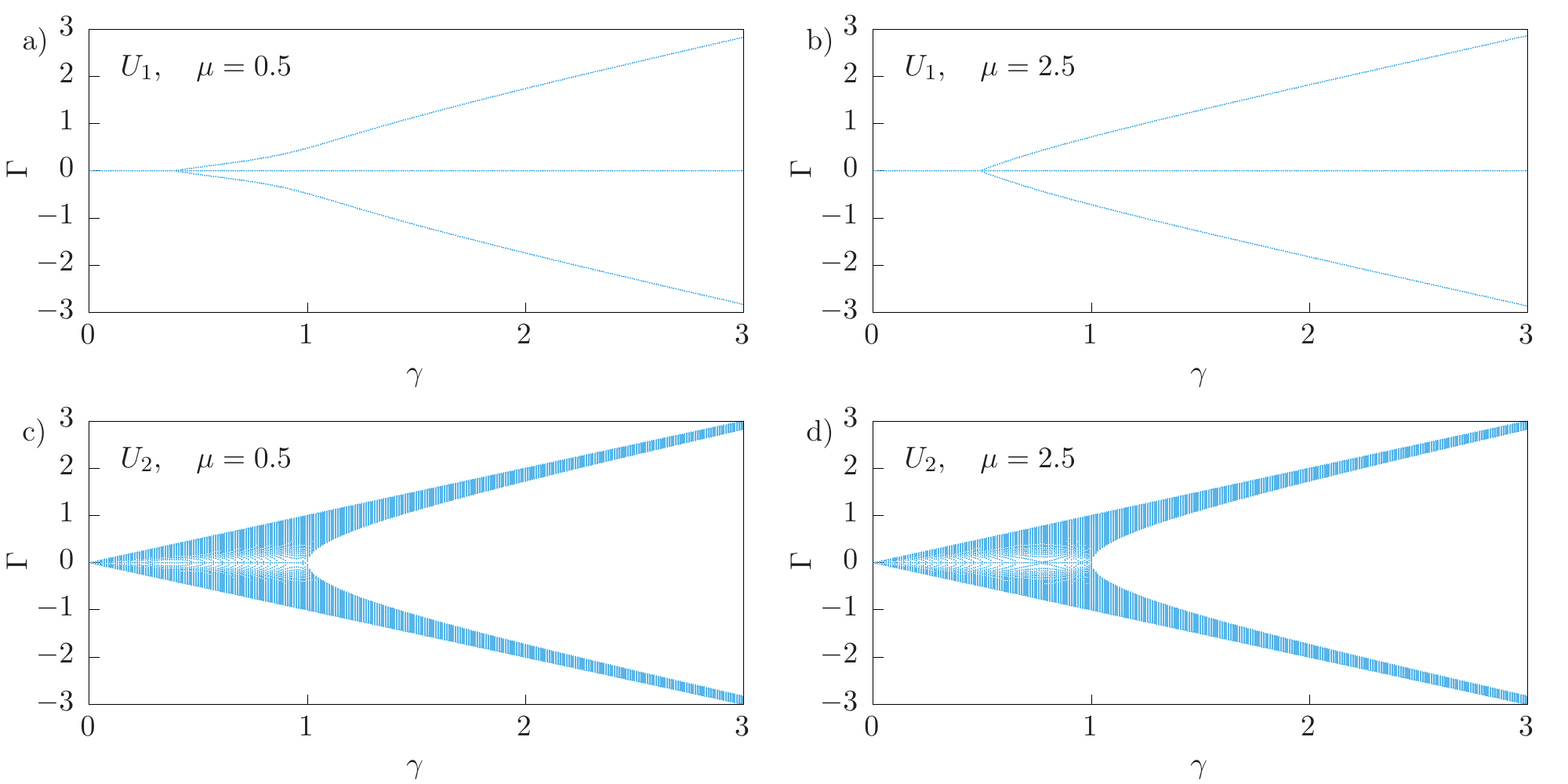}
  \caption{Imaginary parts of all energy eigenvalues for
    the Kitaev chain with length $N=200$ and the parameters $t=1.0$
    and $\delta=t$. The data in a) and b) is calculated with the potential
    $U_1$. In c) and d) $U_2$ is applied.}
  \label{fig:kitaev-pt-phasediagram}
\end{figure*}

For the Kitaev chain with applied potentials $U_1$ and $U_2$ the
imaginary parts of the eigenvalues are shown in
figure~\ref{fig:kitaev-pt-phasediagram}
for two different chemical potentials, where each value represents one of the
topological phases in the isolated case, i.e., $\mu=0.5$ for the TNP and
$\mu=2.5$ for the TTP.  For both potentials there is no obvious difference in
the imaginary part of the energies for the different values of $\mu$. As
in the scenario of the SSH model the potential $U_2$ also exhibits a
critical parameter value $\gamma$ at which no purely real energy eigenvalue
exists. Due to the fact that the chemical potential has no appreciable
influence on the behavior of the imaginary parts of the eigenvalues
the bulk states are responsible for breaking the $\mathcal{PT}$
symmetry. The edge states in the Kitaev model are particle-hole
symmetric and therefore always conserve the $\mathcal{PT}$ symmetry as long as
they exist. Due to a complete $\mathcal{PT}$-symmetry breaking in the case of
the potential $U_2$ the chemical potential capable of hosting edge states is
dependent on the potential strength.

\subsection{Phase diagram for the Kitaev model}

\begin{figure}[tbp]
  \centering
  \includegraphics[width=0.49\textwidth]{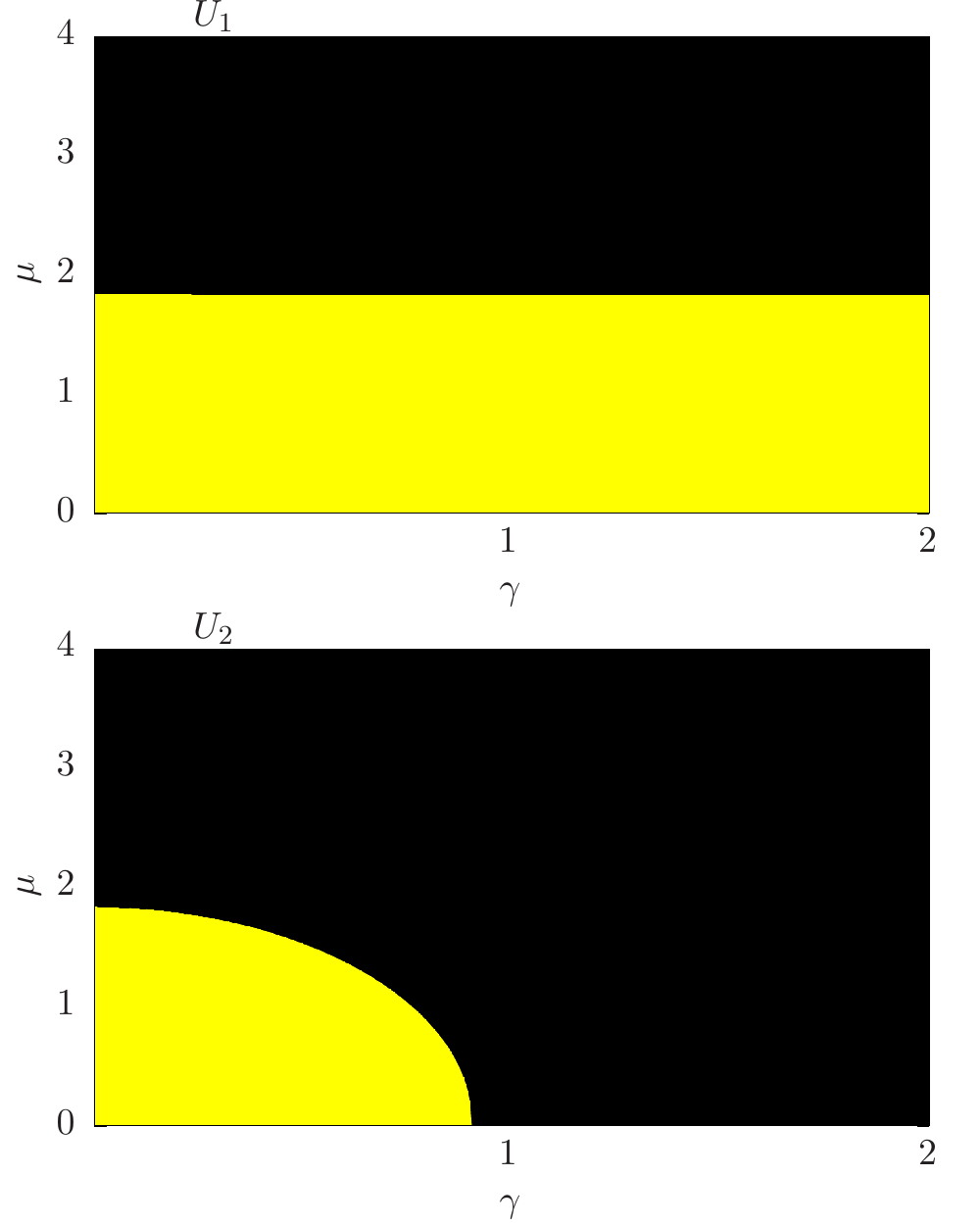}
  \caption{The number of states fulfilling the property
    of Eq.\ \eqref{eq:property-kitaev} for the Kitaev chain with
    $N=200$ lattice sites and the parameters $t=1.0$ and $\delta = t$
    in combination with the potential $U_1$ in a) and $U_2$ in b). The
    color represents the number of states, where yellow (bright)
    stands for the value 2 and black (dark) for the value 0.}
  \label{fig:kitaev-tp-phasediagram}
\end{figure}

Since the potential $U_2$ shows a completely $\mathcal{PT}$-broken
regime when the gain and loss strength is increased, the chemical
potential at which topological edge states can be present has to be a
function of the potential strength, i.e., $\mu(\gamma)$. For each $\mu$
a critical parameter value $\gamma_c$ can be found at which the topological
edge states disappear. The definition of topological edge states in
the sense of the Kitaev chain is the fact that the energy has to
fulfill the property,
\begin{equation}
  \label{eq:property-kitaev}
  E = \mathcal{E} = \Gamma = 0 \; ,
\end{equation}
i.e., it has a vanishing real and imaginary part. To obtain a phase
diagram a constant gain and loss effect $\gamma$ is assumed and for a chemical
potential representing the TTP in the isolated model, e.g., $\mu =4.0$, the
energy is calculated. Then the chemical potential is decreased in steps of
equal size and for each step the number of eigenvalues fulfilling Eq.\
\eqref{eq:property-kitaev} is counted. This is repeated for further values of
$\gamma$. In general, numerical calculations do not supply accurate zero
values, and therefore in this work we measure a numerical zero if the modulus
is smaller than $10^{-8}$. The value selected for zero is in this way not a
critical choice because the energy spectra show a pronounced jump to small
numbers if any topological edge state is present. In
figure~\ref{fig:kitaev-tp-phasediagram}
the number of states, which are in agreement with Eq.\
\eqref{eq:property-kitaev} are counted for the ranges $\mu = 0 \ldots 4$ and
$\gamma = 0 \ldots 2$. By applying the potential $U_2$ one can find values of
$\gamma$ for which no chemical potential supports the existence of edge states.
This is in contrast to the Kitaev chain in the case that electrons gain in
probability at the first site and lose at the last site as described with the
potential $U_1$. In fact the number of edge states does in the latter case not
depend on the value of $\gamma$ and is only limited by the chemical potential
as in the isolated problem. The TNP can host two edge states for $\mu
\lessapprox 2.0$. In total, the Kitaev chain is an example, in which gain and
loss effects can have an influence on the parameter regime with edge states.
However, the two model potentials $U_1$ and $U_2$ show that this behavior of
the TNP depends on the shape of the gain and loss effects, and therefore there
is no general statement on the existence of edge states in the presence of
non-Hermitian potentials in the Kitaev chain.

\section{Summary}
\label{sec:summary}

In conclusion we studied both the Kitaev and the SSH model with two
different $\mathcal{PT}$-symmetric non-Hermitian potentials. Our
investigation of the topological interesting edge states explain why the
topological nontrivial phase in the SSH model shows an instantaneously
$\mathcal{PT}$-broken spectrum for an arbitrary small potential strength
$\gamma$, and that the edge states in the case of the Kitaev chain are
protected from violating $\mathcal{PT}$ symmetry due to an intrinsic
particle-hole symmetry. The important fact is that this symmetry survives in
presence of the gain-loss effect. For the Kitaev chain the number of existing
edge states at a certain value pair $(\mu,\gamma)$ depends on the applied gain
and loss effect, and therefore the range of the topological nontrivial phase
can be a function of the potential strength $\gamma$.

Both results clearly demonstrate that the previously assumed connection
between a spontaneously broken $\mathcal{PT}$ symmetry and the topological
phase \cite{Zhu14a,Wang15} does not exist in general. Only the symmetry
of the individual edge states decides whether their presence has an
influence on the $\mathcal{PT}$ symmetry of the whole system. Their existence
alone does not give a useful answer. It is necessary to always determine the
states and to investigate their probability distribution.

Even though this work can explain the role of the edge states in non-Hermitian
$\mathcal{PT}$-symmetric systems there remain a few open questions. The
imaginary potentials are an effective description of the in- and outfluxes
of the probability amplitude and thus for the interaction with an environment.
Much more realistic is the addition or removal of electrons, which can
be simulated very well in the context of master equations \cite{Breuer02a,%
  Trimborn08a,Dast14a}. It will be interesting to see whether signatures of
the results of this work can be found in the dynamics of the master equation,
and investigations in this direction are under way.
Furthermore, for non-Hermitian operators the often used Berry phase
\cite{Berry84a} or its formulation in the Brillouin zone, i.e.\ the Zak phase
\cite{Zak1989a} need extensions to understand the influence of imaginary
potential contributions. Some extensions exist \cite{Liang2013a,Mandal2016a},
however, they are restricted to special cases. A more general way of
identifying topologically different states would be valuable.
Since similar Bosonic systems with topologically nontrivial states are known
\cite{Grusdt2013a} and in the case of cold atoms much better controllable in
an experiment it seems also worthwhile to extend these studies to Bosons.
This might be in particular interesting since Bosonic many-body systems already
with very simple interactions feature an unusual dynamics such as purity
oscillations in the presence of balanced gain and loss
\cite{Dast2016a,Dast2016b}.

\end{document}